\documentclass[aps,prl,reprint,twocolumn]{revtex4-2}

\usepackage{epsfig}
\usepackage{amssymb}
\usepackage{amsfonts}
\usepackage{amsbsy}
\usepackage[all,v2,cmtip,2cell]{xy}
\usepackage{amsmath}
\usepackage{dsfont}
\usepackage{bbm}
\usepackage{upgreek}
\usepackage{amssymb,amscd}
\usepackage{graphicx}
\usepackage{mathrsfs}
\usepackage{amsmath,amsthm}
\usepackage{slashed}
\usepackage{dsfont}
\usepackage{tikz}
\usepackage[utf8]{inputenc}
\makeatletter
\DeclareFontFamily{OMX}{MnSymbolE}{}
\DeclareSymbolFont{MnLargeSymbols}{OMX}{MnSymbolE}{m}{n}
\SetSymbolFont{MnLargeSymbols}{bold}{OMX}{MnSymbolE}{b}{n}
\DeclareFontShape{OMX}{MnSymbolE}{m}{n}{
    <-6>  MnSymbolE5
   <6-7>  MnSymbolE6
   <7-8>  MnSymbolE7
   <8-9>  MnSymbolE8
   <9-10> MnSymbolE9
  <10-12> MnSymbolE10
  <12->   MnSymbolE12
}{}
\DeclareFontShape{OMX}{MnSymbolE}{b}{n}{
    <-6>  MnSymbolE-Bold5
   <6-7>  MnSymbolE-Bold6
   <7-8>  MnSymbolE-Bold7
   <8-9>  MnSymbolE-Bold8
   <9-10> MnSymbolE-Bold9
  <10-12> MnSymbolE-Bold10
  <12->   MnSymbolE-Bold12
}{}

\let\llangle\@undefined
\let\rrangle\@undefined
\DeclareMathDelimiter{\llangle}{\mathopen}%
                     {MnLargeSymbols}{'164}{MnLargeSymbols}{'164}
\DeclareMathDelimiter{\rrangle}{\mathclose}%
                     {MnLargeSymbols}{'171}{MnLargeSymbols}{'171}
\makeatother

\def\be{ \begin{equation} }
\def\ee{ \end{equation}}

\newcommand{\eq}[1]{\begin{align}\begin{split}#1\end{split}\end{align}}

\def\half{\frac{1}{2}}

\def\one{{\hbox{ 1\kern-.8mm l}}}

\def\CA{{\cal A}}

\def\CB{{\cal B}}

\def\CG {{\cal G}}

\def\CN {{\cal N}}

\def\CP {{\cal P}}

\def\CG {{\cal G}}

\def\CB {{\cal B}}

\def\CS {{\cal S}}
\def\CT {{\cal T}}

\def\IZ{{\mathbb{Z}}}

\def\rmk#1{\bigskip\noindent{\bf Remark} }
\def\cnj#1{\bigskip\noindent{\bf Conjecture:} }

\DeclareMathAlphabet{\mathpzc}{OT1}{pzc}{m}{it}

\def\Tr{ \, \textrm{Tr} \, }

\begin{document}

\title{\boldmath Constraints on Symmetry Preserving Gapped Phases\\
 from Coupling Constant Anomalies}

\author{T.~Daniel Brennan}
\affiliation{Department of Physics, University of California San Diego,\\
 \textit{9500 Gilman Drive, La Jolla CA 92093-0319, USA}}

\begin{abstract}
\noindent In this note, we will characterize  constraints on the possible IR phases of a given QFT by anomalies in the space of coupling constants. We will give conditions under which a coupling constant anomaly cannot be matched by a continuous family of symmetry preserving gapped phases, in which case the theory is either gapless, or exhibits spontaneous symmetry breaking or a phase transition.  We additionally demonstrate examples of theories with coupling constant anomalies which can be matched by a family of symmetry preserving gapped phases without a phase transition and comment on the interpretation of our results for the spontaneous breaking of  ``$(-1)$-form global symmetries.'' 
\end{abstract}
\maketitle



\section{Introduction}
\vspace{-0.2cm}
In recent years, the concept of symmetry has been greatly expanded beyond the action of groups on local operators to include the fusion and linking action of all topological operators  (see \cite{Cordova:2022ruw,Freed:2022qnc,Gaiotto:2014kfa,Schafer-Nameki:2023jdn,Brennan:2023mmt,Bhardwaj:2023kri,Shao:2023gho} and references therein). Likewise our understanding of the possible anomalies allowed in QFTs has also greatly expanded \cite{Kaidi:2023maf,Zhang:2023wlu,Sun:2023xxv,Cordova:2023bja,Antinucci:2023ezl}. 
For example, some anomalies can only be activated in the presence of certain backgrounds such as on non-spin manifolds or in particular flux backgrounds \cite{Brennan:2023vsa,Wang:2018qoy,Brennan:2022tyl,Brennan:2023kpo,Brennan:2023ynm,Anber:2019nze,Anber:2020gig}. These backgrounds are related to symmetries, but are not necessarily activated by topological symmetry operators and hence do not correspond to symmetries themselves, but rather correspond to a sort of ``symmetry structure.'' 
In this paper, we will focus on  one such class of anomalies that are not activated by topological defects called \textit{anomalies in the space of coupling constants} or \textit{coupling constant anomalies} (CC anomalies). 

Consider a family of $d$-dimensional quantum field theories $\CT_\theta$ on $X_d$ which is indexed by a continuous set of exactly marginal couplings $\theta\in \CS$. Let us assume that this family of theories $\CT_\theta$ has a group-like global symmetry $\CG$ which is preserved for all $\theta\in \CS$.

Generically, the set of couplings $\theta\in \CS$ forms a continuous, non-compact space. However, there can be additional isomorphisms $\Lambda$ between the theories $\CT_\theta$ and $\CT_{\theta+\lambda}$ so that $\theta\in \CS/\Lambda$.  
This identification allows us to meaningfully compare the partition functions $Z_\CT[\theta]$, $Z_\CT[\theta+\lambda]$. One possible behavior is that the two partition functions are only identical up to a phase:
\eq{\label{ccMTC}
Z_\CT[\theta]=Z_\CT[\theta+\lambda]\times e^{ i \int_{X_d} \omega_d}
}
where $\omega_d$ is an anomalous phase that is only dependent on background gauge fields for $\CG$. When this phase  is quantized  (i.e. ${\rm exp}\left\{ i k \int \omega_d\right\}$ is only $\CG$-gauge invariant for $k\in \IZ$), there are no local, background gauge invariant counter-terms that can trivialize the phase. In analogy with standard anomalies, 
this phase is referred to as an {anomaly in the space of coupling constants} \cite{Cordova:2019jnf,Cordova:2019uob}. 

Although this ``anomaly'' does not correspond to a symmetry (the interface between $\CT_\theta$, $\CT_{\theta+\lambda}$ is generically not topological), there are still physical consequences of the anomaly. 
One feature is that as with standard anomalies, the CC anomaly   can be described by an $(d+1)$-dimensional SPT phase by anomaly inflow: 
\eq{
\CA=\int_{Y_{d+1}} d\Theta(x)\cup \omega_d
}
where $\partial Y_{d+1}=X_d$ and $\Theta(x)$ is a $(d+1)$-dimensional $\CS/\Lambda$-valued function which takes boundary value $\Theta(x)\big{|}_{x\in X_d}=\theta$. This implies that CC anomalies are scale invariant and must be matched along $\CG$-preserving RG flows.

Note that the framework of CC anomalies encompasses the anomalous phases that arise in the spurion analysis of anomalous 0-form global symmetries that are broken by interactions. As with spurions, if we treat $\theta$ as the vev of some condensed field $\theta(x)$, then  
the CC anomaly implies that the domain wall where $\theta(x)$ shifts $\theta\mapsto \theta+\lambda$ will carry a world volume anomaly given by $\omega_d$. 
See \cite{Cordova:2019jnf,Cordova:2019uob,Choi:2022odr,Debray:2023ior} for more discussion and applications. \\

One illustrative example of a family of theories with an anomaly in the space of coupling constants is $4d$ $SU(N)$ Yang-Mills theory. Despite being non-trivially interacting, it  admits an exactly marginal coupling:
\eq{
S_\theta=...+\frac{i\theta}{8\pi^2}\int_{X_d} \Tr[F\wedge F]~. 
}
It is well known that the coupling $\theta$ is $2\pi$-periodic  
due to the fact that the instanton number is quantized:
\eq{
\int_{X_d} \frac{\Tr[F\wedge F]}{8\pi^2}\in \IZ~. 
}
However,  shifting $\theta\mapsto \theta+2\pi$ is non-trivial due to the Witten effect \cite{Witten:1979ey} which implies that the identification 
$\CT_\theta\sim \CT_{\theta+2\pi}$
involves a redefinition of the electro-magnetic charge lattice  \cite{Aharony:2013hda}. 
 
Yang-Mills also has a $\IZ_N^{(1)}$  1-form center symmetry. If we couple the $\IZ_N^{(1)}$ to a background gauge field $B_2\in H^2(X_d;\IZ_N)$, the instanton number becomes fractional:
\eq{
\int \frac{\Tr[F\wedge F]}{8\pi^2}=\frac{N-1}{2N}\int \CP(B_2)~{\rm mod}_\IZ
}
where $\CP(B_2)\in H^4(X_d;\IZ_{2N})$ is the Pontryagin square of $B_2$ \cite{Witten:2000nv,Aharony:2013hda}. 
Thus, the partition functions obey  \cite{Gaiotto:2017yup}:  
\eq{
Z_\CT[B_2;\theta+2\pi ]=Z_\CT[B_2;\theta]\times e^{\frac{\pi i (N-1)}{N}\int \CP(B_2)}~.
}
Here the partition functions of $\CT_\theta,\CT_{\theta+2\pi}$ are related by a quantized phase that is only dependent on the background gauge fields and hence describes a coupling constant anomaly between $\theta$ and the $\IZ_N^{(1)}$ global symmetry which can be described by the SPT phase \cite{Cordova:2019uob}:
\eq{
\CA =i\frac{N-1}{2 N} \int d\theta\cup \CP(B_2) 
} 

This CC anomaly also implies that if
 we allow $\theta$ to be space-time dependent, then the domain wall where $\theta\mapsto \theta+2\pi$ will carry a world-volume anomaly \cite{Cordova:2019uob,Gaiotto:2017tne,Anber:2020xfk}:
\eq{
\CA_{\rm w.v.}=\frac{\pi i (N-1)}{N}\int \CP(B_2)
}
which indicates that the $\IZ_N^{(1)}$ global symmetry is spontaneously broken on the domain wall where the theory must deconfine \cite{Gaiotto:2017tne,Anber:2020xfk}. This can be physically interpreted by noting that as $\theta$ winds through $\theta+2\pi$ inside the defect, it sweeps through $\theta=\pi$ where the theory cannot be trivially gapped due to a mixed $\IZ_N^{(1)}$--time-reversal anomaly \cite{Gaiotto:2017yup,Gaiotto:2017tne,Cordova:2019jnf,Cordova:2019uob}. This is a well known feature that appears for example in axion-Yang-Mills theory where $\theta$ is replaced by a dynamical axion whose domain wall carries a world volume anomaly from UV chiral modes \cite{Callan:1984sa,Jackiw:1981ee,Nielsen:1976hs,Ansourian:1977qe,Kiskis:1977vh,Weinberg:1981eu,Jackiw:1981ee,Callias:1977kg,Jackiw:1975fn,Brennan:2023kpw}.

\vspace{-0.4cm}

\section{Constraints from CC Anomalies}
\vspace{-0.2cm}
In this paper, we will more precisely derive the constraints on RG flows that are imposed by CC anomalies. Consider a 1-parameter family of $d$-dimensional QFTs $\CT_\theta$ indexed by $\theta\in S^1$  on a manifold $X_d^{(p)}=S^{p}\times S^{d-p}$ with a discrete group-like global symmetry 
\eq{
\CG=G^{(p-1)}\times G^{(d-p-1)}~. 
}
For a family of (non-anomalous) $\CG $-preserving theories, there exists a choice of scheme such that 
\eq{\label{Ginvar}
Z_\CT[A+d\lambda;\theta]=Z_\CT[A;\theta]
\quad, \quad \forall \theta\in S^1 }
where $A$ represents the background gauge fields for $\CG$. 
Let us take such a family of theories with a CC anomaly of the form 
\eq{
\CA=i\int \frac{d\theta}{2\pi}\cup \omega_d(A)
}
where 
$\omega_d\in H^d(B\CG;U(1))$ is quantized (${\rm exp}\left\{ik \int \omega_d\right\}$ is only $\CG$-gauge invariant for $k\in \IZ$)  that describes the anomalous phase in the transformation of the partition function under the periodic shift of $\theta$ in \eqref{ccMTC} \footnote{Here, $B\CG$ is the classifying space for the symmetry group $\CG=G^{(p-1)}\times G^{(d-p-1)}$ which is given by: $B\CG:=B^pG^{(p-1)}\times B^{d-p}G^{(d-p-1)}$.}.


Here we will show that if the anomalous phase can be activated on $X_d^{(p)}$, then the coupling constant anomaly can only be matched in the IR by:
\begin{enumerate}
\item A family of gapless theories (i.e. CFTs)
\item A continuous family of gapped theories (i.e. TQFTs) with spontaneously broken $\CG$-symmetry 
\item A discontinuous family of gapped theories -- i.e. a collection of gapped phases with a phase transition
\end{enumerate}
\noindent Here we restrict to unitary theories and gapped phases which are semi-simple and finite \footnote{For an example of a non-finite TQFT, see \cite{Brennan:2024fgj,Bonetti:2024cjk,Antinucci:2024zjp}.}. 

This implies that a family of QFTs with a coupling constant anomaly  cannot flow to a continuous family of $\CG$-preserving  gapped phases (without phase transition) in the IR if the anomalous phase can be activated on $X_d^{(p)}$. 


\vspace{0.2cm}

\noindent \textbf{Comment on SSB of (--1)-form Symmetries:} The identification $\CT_\theta\sim \CT_{\theta+\lambda}$ is also sometimes referred to  as a ``$(-1)$-form global symmetry.''  There has been much investigation into the consequences of this ``symmetry'' and its spontaneous breaking (SSB) \cite{Aloni:2024jpb,Vandermeulen:2022edk,Vandermeulen:2023smx,Sharpe:2022ene,Santilli:2024dyz}. 

One way we can try to understand the SSB of $(-1)$-form global symmetries is by comparing our results on CC anomalies with known results about anomalies of discrete global symmetries. 
For anomalies involving discrete symmetries, it is known that the anomaly can be matched by spontaneously breaking a global symmetry \cite{Cordova:2019bsd}. The analogous statement we derive here is that a CC anomaly can similarly be matched by a family of gapped phases with a phase transition. 
This suggests that the SSB of a ``$(-1)$-form symmetry'' corresponds to a phase transition in the associated parameter space. We leave a more in-depth analysis for future investigation. 

\vspace{-0.4cm}

\subsection{Derivation of Constraints}
\vspace{-0.2cm}
Let us examine if a family of QFTs with  a  coupling constant anomaly as described above can flow to a continuous family of symmetry preserving gapped phases.  
Here we define a continuous family of gapped phases by a family of TQFTs which have isomorphic Hilbert spaces with a smoothly varying inner product. This implies that the partition function 
$Z_{\rm TQFT}[S^{p+1}\times S^{d-p-1};A;\theta]$ is a continuous function over the parameter space. A continuous family of $\CG$-preserving gapped phases is then a continuous family of gapped phases for which the symmetry $\CG$ is preserved (i.e. not spontaneously or explicitly broken) over the entire family. In particular, a continuous family of gapped phases has no phase transition.

As proved in  \cite{Cordova:2019bsd}, a unitary $\CG$-preserving TQFT obeys:
\eq{
Z_{\rm TQFT}[S^p\times S^{d-p};A]\neq 0~. 
}
Here for technical reasons, we will also restrict ourselves to finite TQFTs which have a finite number of operators/finite dimensional Hilbert spaces on any compact manifold and implies that the partition function is also finite on any closed manifold. In particular: 
\eq{
\Big{|}Z_{\rm TQFT}[S^p\times S^{d-p};A]\Big{|}<\infty~.
} 
Thus, for a continuous family of unitary, $\CG$-preserving, finite TQFTs we can define the phase 
\eq{
\frac{Z_{\rm TQFT}[S^p\times S^{d-p};A;\theta]}{|Z_{\rm TQFT}[S^p\times S^{d-p};A;\theta]|}=e^{i \Phi(\theta;A)}~. 
}
The coupling constant anomaly implies that the phase $\Phi$ takes the form:
\eq{
\Phi(\theta;A)=f(\theta)\int \omega_d(A)+g(\theta)
}
where $f(\theta),g(\theta)$ are semi-periodic functions
\eq{\label{semiperiodic}
f(\theta+2\pi)=f(\theta)+1~, ~  g(\theta+2\pi)=g(\theta)+2\pi n 
}
with $n\in \IZ$. For a continuous family of theories, the continuity of the partition function additionally implies that $f(\theta)$ and $g(\theta)$ are also continuous. 
 
However, since we have assumed the theory is $\CG$-preserving, the property \eqref{Ginvar} together with the fact that $\omega_d(A)$ is quantized, requires that $f(\theta)$ only take integer values. Since no continuous function can be both semi-periodic \eqref{semiperiodic} and take only integer values, there is no unitary, finite, $\CG$-preserving, continuous family of gapped phases that can match the CC anomaly. 

Our assumptions can be violated if 1.) the theory is gapless, 2.) if  $\CG$ is spontaneously broken, or 3.)  the partition function jumps discontinuously (for example if $f(\theta)$ is a step function) -- which would signal a discontinuous family of theories with a phase transition. Note that the scenario where $\CG$ is not uniformly broken over $\theta\in S^1$ which corresponds to the case with a phase transition. 
This   classifies the IR phases that can occur when the anomalous phase is non-trivial on $X_d^{(p)}$ and proves our result.

\vspace{-0.4cm}

\section{Example: $4d$ $SU(N)$ Yang-Mills} 
\vspace{-0.2cm}
Let us again demonstrate our results in the example of $4d$ $SU(N)$ Yang Mills with a $\theta$ angle. This family of theories has a $\IZ_N^{(1)}$ symmetry as well as a time reversal symmetry $\CT$ at $\theta=0,\pi$. 

As discussed above, this theory has the CC anomaly 
\eq{\label{CCAnom4dYM}
\CA_{cc}=i\frac{N-1}{2 N} \int d\theta\cup \CP(B_2) 
}
where $B_2$ is the background gauge field for $\IZ_N^{(1)}$ 
so that the anomalous phase $\omega_d(B_2)=\frac{\pi(N-1)}{N}\int \CP(B_2)$ can be activated on $X_4=S^2\times \tilde{S}^2$ \cite{Cordova:2019uob}:
\eq{
e^{i\int_{X^4} \omega_d(B_2)}=e^{\frac{2\pi i}{N}}~,\quad B_2=[S^2]^\vee+[\tilde{S}^2]^\vee~.
}
Additionally, the $\CT$-symmetry at $\theta=0,\pi$ has a mixed anomaly with $\IZ_N^{(1)}$ at $\theta=\pi$:
\eq{\label{CTZNanomaly}
\CA=\frac{\pi i(N-1)}{ N}\int w_1\cup \CP(B_2)~.
}

This family of theories is believed to be matched in the IR by a discontinuous family of $\IZ_N^{(1)}$-preserving trivially gapped phases with a phase transition at $\theta=\pi$. At $\theta=\pi$, the theory is believed to be a non-trivially gapped phase where $\CT$ is spontaneously broken due to results from large $N$, $\CN=1$ SYM, and symmetry enforced gaplessness \cite{Gaiotto:2017yup,Witten:1998uka,Cordova:2019jnf,Cordova:2019bsd}. This conjecture is consistent with our results and all known symmetries and anomalies. 

According to our results, the symmetries and anomalies of this theory may also be matched by a continuous family of gapped phases without a phase transition that spontaneously breaks the $\IZ_N^{(1)}$ center symmetry. Indeed, such a family of gapped phases is described by the action
\eq{
S_\theta=&\frac{iN}{2\pi}\int (da_1-\CB_2)\wedge b_2\\
&+\frac{i N(N-1) \theta}{8\pi^2}\int (da_1-\CB_2)\wedge (da_1-\CB_2)
}
where $a_1,b_2$ are $U(1)$ gauge fields of degree 1,2 respectively, $\CB_2=\frac{2\pi}{N}B_2$ is an integral lift of $B_2$, and $a_1$ shifts under the choice of lift of $B_2$ so that $da_1-\CB_2$ is invariant \footnote{Different choices of integral lift of $B_2\in H^2(X_d;\IZ_N)$ are related by $\CB_2\mapsto \CB_2+2\pi \Lambda_2$ where $\Lambda_2\in H^2(X_d;\IZ)$. }. 
This theory has a (spontaneously broken) $\IZ_N^{(1)}$ global symmetry (generated by $b_2$-Wilson surfaces), which is identified with the UV $\IZ_N^{(1)}$ global symmetry, and time reversal symmetry at $\theta=\pi$. Because of this identification, the theory reproduces the CC anomaly  \eqref{CCAnom4dYM} and the mixed $\CT$-$\IZ_N^{(1)}$ anomaly  \eqref{CTZNanomaly} of $SU(N)$ Yang-Mills. 
 This phase can be reached by coupling $SU(N)$ Yang-Mills to multiple adjoint scalar fields which condense along  orthogonal directions in $\mathfrak{su}(N)$. 

This family of gapped phases also describes the IR limit of a family of scalar QED models with action
\eq{
S=&\int |D\Phi|^2+\frac{1}{2g^2}F\wedge \ast F-V(\Phi)+\frac{iK\theta}{8\pi^2}  F\wedge F
}
where $K=N(N-1)$, $\Phi$ is a complex scalar field of charge $N$, and $V(\Phi)$ is potential with minimum at $|\Phi|^2\neq 0$. 

\vspace{-0.4cm}
\section{Matching CC Anomalies in Continuous Families of $\CG$-Preserving TQFTs}
\vspace{-0.2cm}
We can also demonstrate a family of theories that have a coupling constant anomaly which can be matched by a continuous family of $\CG$-preserving gapped phases -- i.e. a family of $\CG$-preserving TQFTs without phase transition. 

Consider the families of $6d$ and $8d$ $SU(N)$ Yang-Mills theories with a $\theta$ angle:
\eq{
S_{6d}=\int \frac{1}{2g^2}\Tr[F\wedge \ast F]+i\theta_{6d}\int ch_3(F)\\
S_{8d}=\int \frac{1}{2g^2}\Tr[F\wedge \ast F]+i\theta_{8d}\int ch_4(F)
} 
where $ch_n(F)$ is the $n^{th}$ Chern character 
\eq{
ch_n(F)=\frac{1}{n!(2\pi)^n}\Tr\underbrace{\left[F\wedge...\wedge F\right]}_{n-times}~.
}
These theories have a $\IZ_N^{(1)}$ 1-form center symmetry as in $4d$ $SU(N)$ Yang-Mills above. 
Additionally, the $6d$ and $8d$ theories have a $U(1)^{(1)}$ and $U(1)^{(1)}\times U(1)^{(3)}$ global symmetries that couples to $ch_2(F)$ and $ch_{3}(F),ch_2(F)$ respectively. 
Both of these interacting theories have dimensionful gauge couplings and flow to a free theory in the IR. However, they both have CC anomalies, that constrain all non-anomalous $\IZ_N^{(1)}$-preserving RG flows.

Using the standard decomposition of the Chern character into integral Chern classes:
\eq{
ch_3&=\half c_3-\half c_1 c_2+\frac{1}{6}c_1^3\\
ch_4&=-\frac{1}{6}c_4+\frac{1}{12}c_2^2+\frac{1}{6}c_3c_1-\frac{1}{6}c_2c_1^2+\frac{1}{24}c_1^4 
}
we see  for an $SU(N)$ bundle (which has $c_1=0$) that the respective Chern characters obey the quantization 
\eq{
\oint ch_3(F_{SU(N)})\in \half \IZ~, \quad \oint ch_4(F_{SU(N)})\in \frac{1}{12}\IZ
}
on a generic manifold \footnote{Note that on particularly nice manifolds that these quantization conditions can be modified to be less fractional. For example on an almost complex manifold, the third Chern character is integral 
 \cite{Apruzzi:2021mlh,Geiges}. We will not make any such assumptions here.}. 
The $\theta$ angles are thus periodic:
\eq{
\theta_{6d}\sim \theta_{6d}+4\pi\quad, \quad \theta_{8d}\sim \theta_{8d}+24\pi~.
}
When we turn on a background gauge field for the $\IZ_N^{(1)}$ center symmetry, the quantization of the Chern characters is modified and there is a resulting coupling constant anomaly. If we couple to a background $\IZ_N^{(1)}$ gauge field $B_2\in H^2(X_d;\IZ_N)$ which has an integral lift \footnote{Here we will only focus on $\IZ_N^{(1)}$ background for which $B_2\in H^2(X_d;\IZ_N)$ has an integral lift. In general, the fractional part of Chern characters can different when $B_2$ does not have an integer lift.}, then we can compute the fractional part of the Chern class by embedding the $PSU(N)$ bundle into a $U(N)$ bundle as in \cite{Gaiotto:2017yup}. This computation is straightforward and results in 
\eq{
&ch_3=\frac{\hat{c}_3}{2}-\frac{N-2}{2N}B_2\hat{c}_2+\frac{(N-1)(N-2)}{6N^2}B_2^3\\
&ch_4=-\frac{\hat{c}_4}{6}+\frac{\hat{c}_2^2}{12}+\frac{(N-3)}{6N}B_2\hat{c}_3-\frac{K}{6N^2}B_2^2\hat{c}_2\\
&\qquad +\frac{(N-1)K}{24N^3}B_2^4
}
where $K=N^2-3N+3$ and the $\hat{c}_n\in H^{2n}(X_d;\IZ)$ are integral classes which 
are heuristically the parts of the $PSU(N)$ Chern classes $c_n$ that can be lifted to $U(N)$ Chern classes. Note that $\hat{c}_{2,3,4}$ shift under under the choice of integral lift $B_2\mapsto B_2+N \Lambda_2$ for $\Lambda_2\in H^2(X_d;\IZ)$ so that the entire expression for $ch_{3,4}$ is invariant  $B_2$. 

Since the classes $\hat{c}_{2,3,4}$ are summed over in the path integral, we will restrict our attention to shifts 
\eq{\theta_{6d}\sim \theta_{6d}+4\pi N\quad, \quad \theta_{8d}\sim \theta_{8d}+24\pi N^2
}
so that the partition function shifts by a quantized phase involving only $B_2$, signaling a CC anomaly \footnote{Note that alternatively, the phase proportional to $\hat{c}_2$ for $6d$ Yang-Mills and the terms $\hat{c}_{2,3}$ for $8d$ Yang-Mills can be additionally removed by turning on backgrounds for the $U(1)^{(1)}$ and $U(1)^{(1)}\times U(1)^{(3)}$ respectively. }. 

If we will rescale the $\theta$-angles 
\eq{
\Theta_{6d}=2N \theta_{6d}~, \quad \Theta_{8d}=12N^2 \theta_{8d}~, \quad \Theta\sim \Theta+2\pi }
the coupling constant anomalies are given by
\eq{\label{6d8dCC}
\CA_{6d}&=i\frac{(N-1)(N-2)}{3N}\int d\Theta_{6d}\cup B_2^3\\
\CA_{8d}&=i \frac{(N-1)(N^2-3N+3)}{2N}\int d\Theta_{8d}\cup B_2^4
}

Notice that these examples evade our classification due to the fact that for all choices of $X_d^{(q)}=S^{q}\times S^{d-q}$ and $B_2\in H^2(X_d;\IZ_N)$ the anomalous phases are trivial.

Any family of theories that carries the CC  anomaly $\CA_{8d}$ involving a $\IZ_N^{(1)}$ symmetry as above can be matched by a family of $\IZ_N^{(1)}$-preserving  gapped phases given by
\eq{\label{Gpresphase}
S=&\frac{iN}{2\pi}\int \left(db_3 -\frac{N}{2\pi}\CB_2^2\right)\wedge a_4
\\
&+\frac{i \theta\kappa N}{8\pi^2}\int \left(db_3-\frac{N}{2\pi}\CB_2^2\right)\wedge \left(db_3-\frac{N}{2\pi}\CB_2^2\right)
}
where $b_3,a_4$ are $U(1)$ gauge fields of degree 3,4 respectively, $\kappa=(N-1)(N^2-3N+3)$, and $\CB_2=\frac{2\pi}{N}B_2$ is an integral lift of $B_2$ where $b_3$ shifts so that $db_3-\frac{N}{2\pi}\CB_2^2$ is invariant under the choice of lift of $B_2$. It is unknown if there is a deformation of $8d$ Yang Mills that flows to this gapped phase. 

On the other hand, although our results provide no constraints on the IR phase of $6d$ Yang-Mills, it is unclear how to construct an analogous $\IZ_N^{(1)}$-preserving family of TQFTs that matches the CC anomaly $\CA_{6d}$ in \eqref{6d8dCC}. 

Additionally, as mentioned above, both $6d$ and $8d$ Yang-Mills have additional  global symmetries -- $U(1)^{(1)}$ and $U(1)^{(1)}\times U(1)^{(3)}$ respectively (which couple to the Chern characters $ch_{2,3}(F)$ as appropriate) -- as well as $\CT$-symmetry. These symmetries all have mixed anomalies with $\IZ_N^{(1)}$ due to the fractional part of the Chern characters.  For $8d$ Yang-Mills, the mixed $\CT$-$\IZ_N^{(1)}$ anomaly is matched by the family of TQFTs above. However it is unclear if anomalies involving continuous symmetries can be matched by a symmetry preserving gapped phase. If such anomalies were to obstruct a $\IZ_N^{(1)}\times U(1)^{(1)}\times U(1)^{(3)}$-preserving gapped phase for $8d$ Yang-Mills, this would imply that any deformation that causes the UV theory to flow to the gapped phase in \eqref{Gpresphase} (if any such deformation exists) must necessarily break $U(1)^{(1)}\times U(1)^{(3)}$.\\ 

The fact that $\CA_{8d}$ can be matched by a family of $\IZ_N^{(1)}$-preserving gapped phases and that the CC anomaly of $4d$ $SU(N)$ Yang-Mills can be matched by a continuous family of gapped phases with $\IZ_N^{(1)}$ SSB 
demonstrates that coupling constant anomalies do not always imply the existence of a phase transition in the IR.

\vspace{-0.4cm}

\section*{Acknowledgements}
\vspace{-0.2cm}
The authors would like to thank Ken Intriligator,  Po-Shen Hsin, Kantaro Ohmori, Anuj Apte, Sungwoo Hong, and Clay C\'ordova  for helpful discussions and related collaborations. 
TDB is supported by Simons Foundation award 568420 (Simons
Investigator) and award 888994 (The Simons Collaboration on Global Categorical Symmetries). 
 
\bibliographystyle{utphys}
\bibliography{CCgaplessnessBib}

\end{document}